\documentclass[prl,fleqn,showpacs,aps,twocolumn,10pt]{revtex4-1}
\usepackage{graphicx}
\usepackage{amsmath}
\usepackage{subfigure}
\usepackage{epsfig}
\usepackage{setspace}
\usepackage{xcolor}
\begin{document}

\title{Collective Photon Assisted Dressing of Atomic Levels by the 
number $N$ of Correlated Atoms}

%\title{Collective Atom Levels Dressing by Photon Assisted N Correlated Atoms }

%\title{Dressing of Two Level Correlated Atoms by the Number of Atoms\\Many body photonic cavity quantum correlation: case study two atoms two levels system\\
%Cavity photonic induced two quantum systems correlations via excitation exchange operator}

\author{Reuben Shuker}
\email{shuker@bgu.ac.il}
\affiliation{Physics Department, Ben Gurion University of the Negev, P.O.Box 653, Beer Sheva, 8410501, Israel}

\author{Gennady~A. Koganov}
\email{quant@bgu.ac.il}
\affiliation{Physics Department, Ben Gurion University of the Negev, P.O.Box 653, Beer Sheva, 8410501, Israel}

\begin{abstract}
Enhancement of the sensitivities of optical magnetometers, atomic clocks and atom interferometers and other quantum metrology devices requires introducing new physical processes to improve on their present  achievements.  
Many body collective correlations among the atoms, spins or, in general, quantum systems may prove to be a suitable method. As these correlations introduce interference terms in the intensity of the scattering amplitudes, they may enhance the signal as $N(N-1)$ for N correlated quantum systems. These  correlations enhance the signal to noise ratio by a factor of $N^2$ and contribute to better sensitivity in quantum metrology. Moreover atomic correlation may provide quantum noise limit, Heisenberg limit. In the present communication excitation exchange induced by photons in a cavity between two atoms is calculated and clearly exhibits correlation and collective effects. A novel operator is introduced that expresses photon-induced excitation exchange that takes in account energy conservation, $V_{ij}=\hat{a}^\dag\sigma_i\sigma_j^\dag\hat{a}$, $\sigma_i=\left|g\right\rangle_{i}\left\langle e\right|_{i}$ is lowering operator of \textit{i-th} atom, and $\hat{a}^\dag,\hat{a}$ are photon creation and annihilation operators. Here i and j stand for two atoms. This operator describes real or virtual photon assisted dipole-dipole interaction. Moreover, it conserves the total number of excitations in the joint em field and the quantum system. Experimental challenges are suggested.
%Interatomic interactions play ...
%Pair atomic interactions in magnetometry, atomic clocks...
\end{abstract}

\maketitle

%\section{Introduction}

Impressive achievements have been obtained in sensitivity of atomic clocks and optical magnetometers that practically arrive at the theoretical limit \cite{Romalis-subfemto,Clocks-McGrew,SERF-Review}. In both the signals are linear in the number $N$ of active atoms. Enhancement of these sensitivities requires introducing new physical processes that may improve on these achievements. 
Many body collective correlations among the atoms or in general quantum systems, may prove the right method \cite{Kimble2016,Ritsch,Zoller,photonic,Javanainen}. When $N$ quantum systems (atoms) interact with electromagnetic field the response intensity of the system is $\left|\sum_{i=1}^{N}{a_i}\right|^2$, where $a_i$ is the response amplitude of a single atom. This can be written as $\sum_{i=1}^{N}{\left|a_i\right|^2}+\sum_{i\neq j}^{N}{\left|a_i a_j\right|}$. The second term stems from pair correlations. If the atoms were statistically independent, this correlation function is equal to zero, and the system's response intensity is merely equal to response intensity of an individual atom multiplied by $N$, as described by the first term. As these correlations introduce interference terms in the intensity of the amplitudes, they may enhance the signal up to $N^2$ for N correlated quantum systems, in case of constructive interference, as in super-radiance \cite{Haroche}. In the case of destructive interference the interference term may quench spontaneous emission and quantum noise \cite{Kimble2016}. Both of these cases, if correlation is obtained, result in better signal to noise ratio and may contribute to better sensitivity in atomic clocks, optical magnetometers, atomic interferometry \cite{Narducci} and other quantum processes, such as measurements, computing and communications. Atomic correlations may also reduce phase quantum noise and achieve the Heisenberg limit \cite{Mitchel-subprojection,HisenLimit}.\\

In this communication we introduce a physics of generating correlations among quantum systems, for example two atoms, by cavity photonic excitation exchange. We present a case study of $N$ two-levels systems interacting with cavity photons in quest of quantum correlations. To take account for pair atomic correlations we define a collective excitation exchange operator $V$as follows:

\begin{equation}\label{V1}
\hat{V}=\Omega_c\sum_{i\neq j}{\hat{a}^\dag\sigma_i\sigma_j^\dag\hat{a}},
\end{equation}

\noindent where $\sigma_i=\left|g\right\rangle_{i}\left\langle e\right|_{i}$ is lowering operator of \textit{i-th} atom, and $\hat{a}^\dag$ and $\hat{a}$ are photon creation and annihilation operators, $\Omega_c$ is the strength of direct interatomic interactions. The main virtue of this operator is that it conserves energy and the total number of excitations in the matter and photonic systems, as a whole. Excitation exchange operators, such as $\sigma_i\sigma_j^\dag$ \cite{antiresonance} or Stokes operator \cite{Mitchell-2014}, are energy non-conserving. This is crucial to make the total Hamiltonian energy conserving.

Our novel operator explicitly presents fully quantum mechanically photon assisted atom correlation while it is absent in other works \cite{antiresonance,Mitchell-2014,superrad-dipole-int}. As will be seen later, we are able to obtain dressing by the number of atoms $N$. Indeed, the operator (\ref{V1}) also correctly generates the necessary spin-spin correlation and their collective behavior treated in Refs.\cite{Mitchell-2014,Mitchel-subprojection}. In fact, spin-spin correlation is generated due to Fermi-Dirac statistics of spins as presented in \cite{Mitchel-subprojection,Mitchell-2014}. However, dressing by the number of atoms $N$ is not generated in these works, in particular in the case where photons are treated semiclassically \cite{antiresonance,superrad-dipole-int}. Also, the virtue of atom correlations may provide the possibility of achieving Heisenberg limit \cite{Mitchel-subprojection,HisenLimit} of quantum noise $1/N$, rather than standard quantum noise limit of $1/\sqrt{N}$.\\

%\section{Model}

We introduce a general system of $N$ two-level atoms interacting with a resonant single mode of electromagnetic field. The system Hamiltonian in the interaction picture and the rotating wave approximation reads

\begin{equation}\label{hamilt}
	H=\sum_{i=1}^{N}{(g\sigma_i}\hat{a}^\dag+H.c.)+V,
\end{equation}

%\begin{equation}\label{hamilt}
%	H=\sum_{i=1}^{N}{(g\sigma_i}\hat{a}^\dag+H.c.)+\Omega_c \sum_{i\neq j}{\hat{a}^\dag\sigma_i\sigma_j^\dag\hat{a}},
%\end{equation}

\noindent where g is a coupling constant, and the collective operator $V$ is defined in Eq. (\ref{V1}). The first term in (\ref{hamilt}) describes the standard interaction of the atoms with the field, while the second term is responsible for field assisted pair atomic correlations, such as dipole-dipole and spin-spin interaction. It generates collective term proportional to $N(N-1)$ as a result of the interference term and the related correlation. \\

%\section{Eigenvalues and dressed states}

In the following we present the collective effect and dependence of eigenvalues and dressed states on the number of atoms $N$. Calculation of the eigenvalues of the Hamiltonian (\ref{hamilt}) results in

\begin{equation}\label{eigen}
	E=\frac{1}{2}N[\Omega_c(N-1)n\pm\sqrt{4ng^2+\Omega_c^2n^2(N-1)^2} ],
\end{equation}

\noindent where $n$ is the number of photons inside the cavity. Note that the eigenvalues (\ref{eigen}) manifest collective effect, as seen by asymptotic $N^2$ dependence at large values of $N$:

\begin{align}
E_1\approx \Omega_c n N^2,\label{asymptup}\\
E_2\approx -\frac{g^2}{\Omega_c}+O(\frac{1}{N^2})\label{asymptlow}
\end{align}

\begin{figure}[htbp]
	\centering
		\includegraphics[scale=0.35]{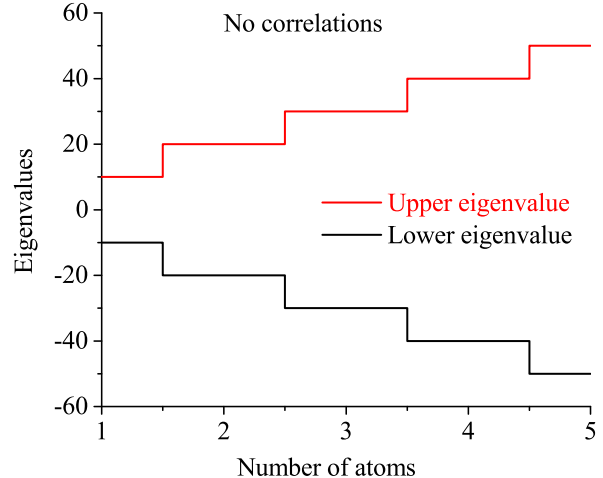}
	\caption{(Color online) Upper and lower Eigenvalues of Hamiltonian (\ref{hamilt}) without correlation term.}\label{eigvalnoCorr}
\end{figure}

 Obviously no collective effect when $N=1$. Indeed, in this case the usual photon splitting effect leading to Autler-Towns doublet \cite{AutlerTownes} is found. For $n=0$ no cavity effect and no splitting. However, even for $n < 1$ cavity and correlation effects are possible \cite{Thompson}. In Fig. \ref{eigvalnoCorr} the eigenvalues (\ref{eigen}) are plotted as a function of the number of atoms $N$ in the absence of correlations $\Omega_c=0$. Both upper and lower eigenvalues change with constant step equal $g\sqrt{n}$ and are symmetric (with respect zero).

\begin{figure}[htp]
	\centering
		\includegraphics[scale=0.35]{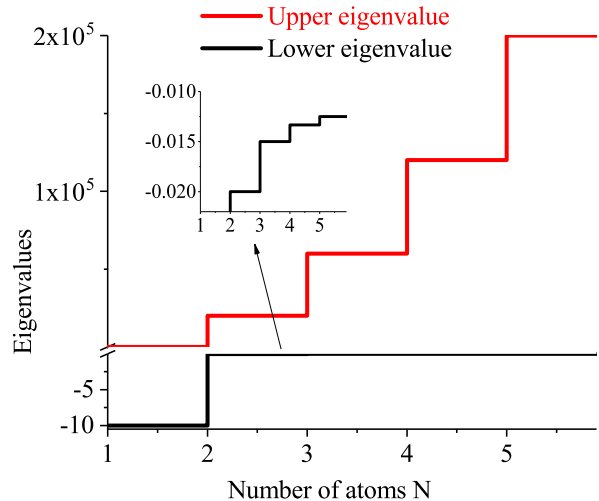}
	\caption{(Color online) Upper and lower eigenvalues of Hamiltonian (\ref{hamilt}) with correlation term.}\label{eigvalCorr}
\end{figure}

When correlations are present, the eigenvalues change with $N$ in a different way (see Fig. \ref{eigvalCorr}): the upper eigenvalue grows with the number of atoms with the rate proportional to $N$, i.e. it scales as $N^2$, whereas the lower eigenvalue does not decrease, as in the previous case, but grows slowly with the rate proportional to $1/N^2$ and asymptotically tends to negative limit value. These can be seen from the asymptotic expressions (\ref{asymptup}) and (\ref{asymptlow}). Both eigenvalues change with varying stair step, which can also be seen from the asymptotic expressions for the stair step values $\Delta E=E(N+1)-E(N)$ at large $N$.

\begin{align}
\Delta E^{(+)}=2 \Omega_cnN+\textit{O}\left(\frac{1}{N^2}\right) \label{E+}\\
\Delta E^{(-)}=\frac{g^2}{\Omega_c}\frac{1}{N^2}+\textit{O}\left(\frac{1}{N^3}\right)\label{E-}
\end{align}

As seen from comparison of Figs. \ref{eigvalnoCorr} and \ref{eigvalCorr} the atomic correlations introduce symmetry breaking.

\begin{figure}[htp]
	\centering
		\includegraphics[scale=0.35]{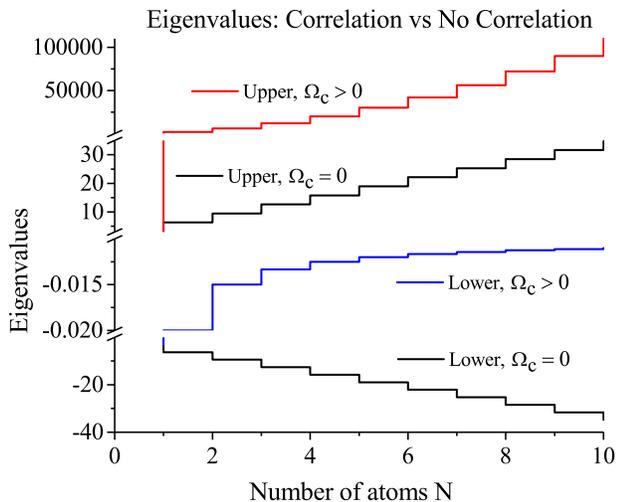}
	\caption{(Color online) Eigenvalues of the Hamiltonian (\ref{hamilt}) with positive correlations ($\Omega_c>0$) and without correlations (black lines).}\label{eigval-Corr-NoCorr}
\end{figure}

\begin{figure}[htp]
	\centering
		\includegraphics[scale=0.35]{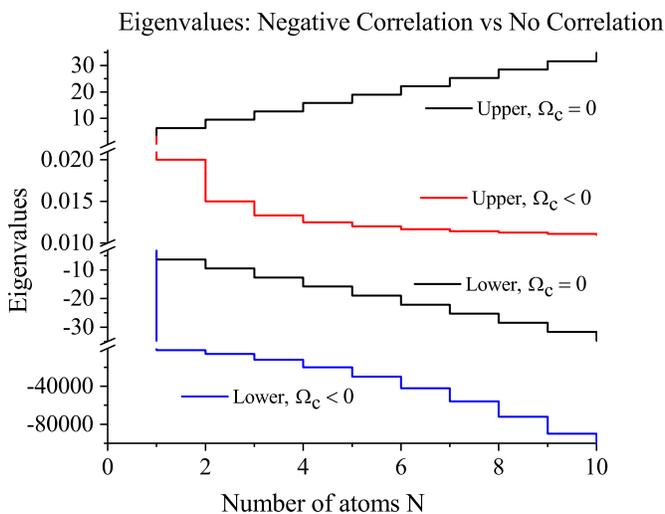}
	\caption{(Color online) Eigenvalues of the Hamiltonian (\ref{hamilt}) with negative correlations ($\Omega_c<0$) and without (black lines) correlations.}\label{eigval-AntiCorr-NoCorr}
\end{figure}

Figures \ref{eigval-Corr-NoCorr} and \ref{eigval-AntiCorr-NoCorr} show the symmetry breaking for two kinds of correlation, positive correlation with $\Omega_c>0$ (Fig. \ref{eigval-Corr-NoCorr}), and negative correlation with $\Omega_c<0$ (Fig. \ref{eigval-AntiCorr-NoCorr}). In the first case the upper eigenvalue grows as $N^2$ at large $N$, while the lower one asymptotically approaches negative limit value $-g^2/\Omega_C$, as described by the approximate expressions (\ref{asymptup}) and (\ref{asymptlow}), respectively.

In the case of negative correlation, $\Omega_c<0$, symmetry is broken in a different way, somewhat opposite to the case of positive correlation, as seen in Fig. \ref{eigval-AntiCorr-NoCorr}. Here the upper eigenvalue at large number of atoms, $N>>1$, asymptotically approaches positive limit value $g^2/\left|\Omega_c\right|$, while the lower eigenvalue being negative scales as $N^2$. 

\begin{figure}[htp]
	\centering
		\includegraphics[scale=0.35]{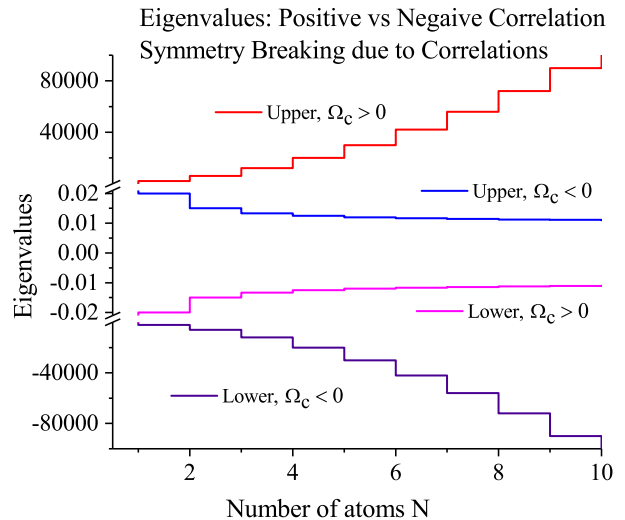}
	\caption{(Color online) Eigenvalues of the Hamiltonian (\ref{hamilt}) with positive ($\Omega_c>0$) and negative correlations ($\Omega_c<0$).}\label{eigval-Corr-AntiCorr}
\end{figure}

In Fig. \ref{eigval-Corr-AntiCorr} eigenvalues in two cases of positive and negative correlations are presented, elucidating symmetry breaking. This is a kind of restoring symmetry, namely, there is an inversion symmetry between positive and negative correlations. 

\begin{figure}[htp]
	\centering
		\includegraphics[scale=0.3]{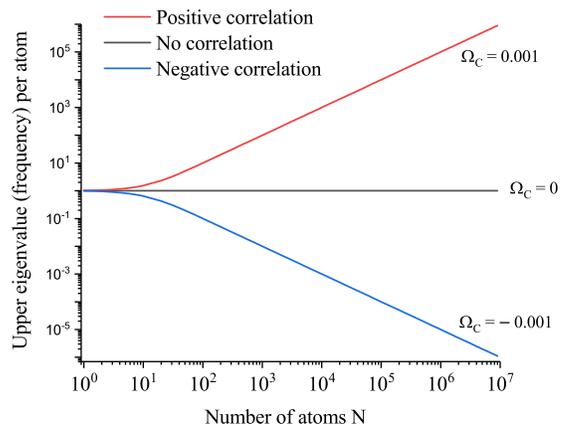}
	\caption{(Color online) Dressed states transition frequency (eigenvalues of the Hamiltonian (\ref{hamilt})) per single atom as a function of the number of atoms $N$ at positive, negative and no correlation cases.\label{eigval-per-Atom}}
\end{figure}

Finally, we present in Fig. \ref{eigval-per-Atom} eigenvalues (actually, these are frequencies of the transitions between the dressed states) per single atom. In the case of non-correlated atoms this frequency does not depend on the number of atoms $N$, as expected. When the atoms are correlated, the dressed transition frequency per atom grows linearly with $N$, both at positive and negative correlation.

In summary, the novel operator describing field-assisted pair correlations among quantum systems, was introduced. A case study of $N$ two-levels systems interacting with cavity photons in quest of quantum correlations has been presented. Calculation of the eigenvalues of the Hamiltonian containing the new term responsible for quantum correlations shows that accounting for such correlations gives rise to breaking the system symmetry. The upper eigenvalue is no longer linear with respect to the number of atoms $N$, as it is when the atoms are independent, but scales as $N^2$ for $N>>1$. It has also been found that when the correlation is negative, the system symmetry breaks in a different way, namely, the lower (negative) eigenvalue of the Hamiltonian scales as $N^2$, while the upper one asymptotically drops to small limit value. The symmetry is restored taking both positive and negative correlation constant.\\

Our novel operator explicitly presents in the Hamiltonian quantum mechanically photon assisted atom correlation while it is absent in other works \cite{antiresonance,Mitchell-2014,superrad-dipole-int}, thus we are able to obtain dressing by the number of atoms $N$. Other virtue of atom correlations may provide the possibility of achieving Heisenberg limit of quantum noise $1/N$, rather than standard quantum noise limit of $1/\sqrt{N}$.

An experimental test for this theory should start with a cold atoms with anti-reflection coating optical cell inserted into a strong coupling regime cavity. A first attempt setup may involve cold Rb atoms to reduce thermal relaxations and extend $T_2$. We suggest to employ lasing without inversion in a cavity with one bi-chromatic mirror, such that a laser at $D_2(D_1)$ can pass and $D_1(D_2)$ is reflected in the cavity. Monitoring the $D_1$ line intensity vs the number density of the cold atoms by controlling the temperature would provide the test for atoms correlation effect. Successful result may pave the way for atoms correlation. We believe that atoms correlation may prove robust effect even at hot vapor case.

\section*{Acknowledgment}

We acknowledge the support of Office of Naval Research Grant No. ONRG GRANT- NICOP-N62909-19-1-2030.

\bibliographystyle{unsrt}
\bibliography{Koganov}

\end{document}